\documentclass[a4paper]{article}
\usepackage{amsfonts}
\usepackage{amssymb}
\usepackage[margin=2cm]{geometry}
\usepackage[dvips]{graphicx}

\usepackage[T1]{fontenc}
\usepackage[latin2]{inputenc}

\usepackage{amsmath}
\usepackage{bbm}

\begin{document}


\title{A simple test of quantumness for a single system}

\author{Robert Alicki \\
  {\small
Institute of Theoretical Physics and Astrophysics, University of
Gda\'nsk,  Wita Stwosza 57, PL 80-952 Gda\'nsk, Poland
}\\
Nicholas Van Ryn\\
  {\small
School Of Physics, Quantum Research Group, University of
KwaZulu-Natal,}\\{\small Westville Campus, Private Bag x54001,
Durban, South Africa.}\\ }

\date{\today}
\maketitle

\begin{abstract}
We propose a simple test of quantumness which can decide whether for the given set of accessible experimental data the classical model is insufficient. Take two observables $ A,B$ such that for any state  $\psi$ their mean values
satisfy $0\leq \langle\psi|A|\psi\rangle\leq \langle\psi|B|\psi\rangle\leq 1$. If there exists a state $\phi$ such that the second moments
fulfill the inequality $\langle\phi|A^2|\phi\rangle >\langle\phi|B^2|\phi\rangle $ then the system cannot be described
by the classical probabilistic scheme. An example of an optimal triple $(A,B,\phi)$ in the case of a qubit is given.

\end{abstract}
Although we are confident that the proper theory describing all
physical phenomena is the quantum theory, there are many situations
where the classical description in terms of functions and
probability distributions over a suitable "phase-space" is
sufficient. In particular, the systems consisting of a large number
of particles and/or emerging in quantum states characterized by
large quantum numbers are supposed to behave classically. The
standard explanation of this fact refers to the stability properties
of quantum states with respect to the interaction with an
environment. For large quantum systems the interaction with the
environment is so strong that most quantum states rapidly decay
(decohere) and the remaining manifold of experimentally accessible
states can be described by classical models. However, the actual
border between quantum and classical worlds is still a topic of
theoretical debate and experimental efforts \cite{dec}. This
question is particularly important in the field of quantum
information. Useful large scale quantum computations would demand
preservation of some quantum properties for rather large physical
systems, say at least for $10^3$ qubits. Moreover, some promising
implementations of a qubit are based on mesoscopic systems which lie
on the aforementioned border. In particular, the so-called
superconducting qubits are systems composed of $10^8-10^9$ particles
(Cooper pairs). Therefore, it is important to find a simple
operational test of quantumness which could be applied to a single
system implementing a qubit. It is generally believed that the
simplest operational and model independent test of quantumness (or
strictly speaking non-classicality) is based on Bell inequalities
which involve systems composed of at least two parts \cite{bell}.
However, this test is problematic for systems which cannot be
spatially well-separated ("locality loophole").

We propose here a much simpler operational test which can be applied
to a single system and formulated in terms of inequalities between
mean values of properly chosen observables. In contrast to Bell
inequalities this test does not refer to the notion of locality,
associated with the space-time,  but involves only the most
fundamental differences between classical and quantum observables.
It is based on the observation that for any two real functions $f,g$
satisfying
\begin{equation}
 0\leq f(x)\leq g(x)
\label{ine}
\end{equation}
and any probability distribution $\rho(x)$ the following inequality holds
\begin{equation}
\langle f^2\rangle_{\rho} \equiv \int f^2(x)\rho(x)dx \leq \int
g^2(x)\rho(x)dx \equiv \langle g^2\rangle_{\rho} \ . \label{main}
\end{equation}
On the other hand, as illustrated below for the case of a single
qubit, for quantum systems we can always find a pair of observables
$A, B$ such that for all states  $\psi$ their mean values satisfy
$0\leq \langle\psi|A|\psi\rangle\leq \langle\psi|B|\psi\rangle$ but
there exists a state $\phi$ such that for the second moments
$\langle\phi|A^2|\phi\rangle >\langle\phi|B^2|\phi\rangle $.
\par
The above statement follows from the interesting and nontrivial
mathematical result in the theory of $C^*$-algebras \cite{kad},
which is presented below for completeness.
\par
Consider an abstract formalism  where (bounded) observables are
elements of a certain $C^*$-algebra ${\cal A}$, and states are
positive normalized functionals ${\cal A}\ni A\mapsto\langle A\rangle_{\rho}$ with
$\langle A\rangle_{\rho}$ denoting the mean value of an observable $A$ in a state
$\rho$. For all practical purposes we can restrict ourselves to two
extreme cases: the first being a classical model where ${\cal A}$ is
an algebra of functions on a certain "phase-space" with $\langle A\rangle_{\rho}=
\int A(x)\rho(x)dx$, where $\rho(x)$ is some probability distribution, and
the second being a finite quantum model in which ${\cal A}$ is an
algebra of matrices and $\langle A\rangle_{\rho}={\rm Tr}(\rho A)$, where $\rho$ is
some density matrix. For any pair of observables $A,B \in {\cal A}$, the order
relation $A\leq B$ means that $\langle A\rangle_{\rho}\leq \langle B\rangle_{\rho}$ for all states
$\rho$ (in fact it is enough to take all pure states).  Now we can
formulate the following:
\par
{\bf Theorem.} \emph{If the following implication
\begin{equation}
0\leq A\leq B \Longrightarrow A^2 \leq B^2
\label{imp}
\end{equation}
always holds then the algebra ${\cal A}$ is commutative, i.e. isomorphic to the algebra of
continuous functions on a certain compact space.}
\par
As a consequence of the above theorem, for any quantum system there
exists a pair of observables (identified with matrices) $(A,B)$ such
that the eigenvalues of $A, B$ and $B-A$ are nonnegative but the
matrix $B^2-A^2$ possesses at least one negative eigenvalue.
\par
In order to apply our test of quantumness in an experiment, one
should first guess a pair of observables $A,B$ with nonnegative
values of possible outcomes and perform a statistical test of the
inequality  $\langle A\rangle_{\rho} \leq \langle B\rangle_{\rho}$ with as large as possible number
of different, generally mixed, initial states $\rho$. These states should be as
pure as possible, otherwise the quantumness could be not detected. Then one should search amongst these for any
states $\sigma$ satisfying the relation for the second moments
$\langle A^2\rangle_{\sigma} > \langle B^2\rangle_{\sigma}$. If the violation  of the classical
relation (\ref{main}) holds, it means that the system exhibits some
quantum characteristics. Although there are always infinitely many
triples $(A,B,\sigma)$, the effects of external noise acting on the
system and measuring apparatus can easily wash out the deviations
from "classicality". Therefore, instead of a random guess it is
useful to find the examples of such triples which maximally violate
classicality and can be used to optimally design the experimental
setting. This can easily be achieved in the case of a qubit which is
the most important example for quantum information.
\par
We search for a pair of $2\times 2$ matrices $A, B$ and a pure state $\phi$
which satisfy
\begin{equation}
0\leq A\leq B\leq I \ , \langle A^2\rangle_{\phi}\equiv\langle\phi|A^2|\phi\rangle
>\langle B^2\rangle_{\phi}\equiv\langle\phi|B^2|\phi\rangle \ , \label{ex}
\end{equation}
where the observables $A$ and $B$ are normalized in such a way that
their upper bound is the identity. Such a triple is given as an
example as
\begin{equation}
A =\begin{pmatrix}a_1 & \xi\cr
            \xi^* & a_2
\end{pmatrix} ,\
B =\begin{pmatrix} 1 & 0\cr
            0 & b
\end{pmatrix}
            ,\
\phi= \begin{pmatrix} \alpha\cr
            \beta
\end{pmatrix}.
\label{qubit}
\end{equation}
The matrix $B$ is chosen to be diagonal, with the identity as its
upper bound, and this choice of basis can be made since the solution
to this problem is unique up to unitary equivalence. From the upper
bound it can be seen that both eigenvalues of $B$ should be at most
$1$, and to maximize the violation of (\ref{imp}) one of the
eigenvalues is chosen to be fixed at $1$. The usual condition
$|\alpha|^2+|\beta|^2=1$ applies for the parameters of the state
$\phi$. The positivity of the observables $A$, $B$ and $(B-A)$ is
ensured by the requirement that both their diagonal elements and
determinants are positive. These conditions are expressed below as;
\begin{equation}
0\leq b\leq 1,\label{cond1}
\end{equation}
\begin{equation}
0\le a_1\le 1,\label{cond2}
\end{equation}
\begin{equation}
0\le a_1a_2-|\xi|^2,\label{cond3}
\end{equation}
and
\begin{equation}
0\le (1-a_1)(b-a_2)-|\xi|^2.\label{cond4}
\end{equation}
In order for this triple to satisfy equation (\ref{ex}), it is
sufficient that one of the eigenvalues of $(B^2-A^2)$ is found to be
negative while remaining within the constraints listed above.

The eigenvalues of the matrix $(B^2-A^2)$ are found to be
\begin{equation}
\frac{1}{2}\left(b^2+1-a_1^2-a_2^2-2|\xi|^2\pm\sqrt{(b^2-1+a_1^2-a_2^2)^2+4(a_1+a_2)^2|\xi|^2}\right).
\end{equation}
In order to find the most negative value which one of the
eigenvalues can attain, one need only consider one of the two. Since
the square root is always positive, one eigenvalue always remains
greater than the other. A numerical technique is used to calculate
the maximal violation of the inequality (\ref{imp}) since we are
unable to find an exact solution to this optimisation problem. In
finding the optimal set of parameters $a_1$, $a_2$, $b$ and real
$\xi$, it can be seen that the maximal violation of the inequality
(\ref{imp}) arises when the conditions (\ref{cond3}) and
(\ref{cond4}) are equalities rather than inequalities. Using
\begin{equation}
a_1a_2=|\xi|^2,\label{cond3eq}
\end{equation}
and
\begin{equation}
(1-a_1)(b-a_2)=|\xi|^2\label{cond4eq}
\end{equation}
reduces the problem from four unknowns to two, and the triplet can
then be written as
\begin{equation}
A =\begin{pmatrix}a_1 & \sqrt{a_1 a_2}\cr
            \sqrt{a_1 a_2} & a_2
\end{pmatrix} ,\
B =\begin{pmatrix} 1 & 0\cr
            0 & \frac{a_2}{1-a_1}
\end{pmatrix}
            ,\
\phi= \begin{pmatrix} \alpha\cr
            \beta
\end{pmatrix}.
\label{qubit1}
\end{equation}

The parameters which result
in one of the eigenvalues of $(B^2-A^2)$ attaining it's most
negative value, and thus maximally violating inequality (\ref{imp}),
while still remaining within the constraints give us the triplet
\begin{equation}
A =\begin{pmatrix}0.724 & 0.249\cr
            0.249 & 0.0854
\end{pmatrix} ,\
B =\begin{pmatrix} 1 & 0\cr
            0 & 0.309
\end{pmatrix}
            ,\
\phi= \begin{pmatrix} 0.391\cr
            0.920
\end{pmatrix}.
\label{example}
\end{equation}
In this example, the value of
$\langle\phi|B|\phi\rangle-\langle\phi|A|\phi\rangle$ is $0.0528$, a
positive value, whereas it can be seen that
$\langle\phi|B^2|\phi\rangle-\langle\phi|A^2|\phi\rangle=-0.0590$
which clearly demonstrates the quantum nature of this example. The
eigenvectors and corresponding eigenvalues of $A$ using these
parameters is calculated to be
\begin{equation}
A\begin{pmatrix} 0.946\cr
            0.325
\end{pmatrix}=0.809
\begin{pmatrix} 0.946\cr
            0.325
\end{pmatrix}
\end{equation}
and
\begin{equation}
A\begin{pmatrix} -0.325\cr
            0.946
\end{pmatrix}=
\begin{pmatrix} 0\cr
            0
\end{pmatrix}.
\end{equation}

To give a concrete example, one can apply these results to the
polarization of a single photon. Choosing a polarization basis as
$|H\rangle , |V\rangle$ and attributing the values  $1$ to
$|H\rangle$ and $0.309$ to $|V\rangle$ we obtain the observable $B$.
The observable $A$ corresponds to a rotated polarization basis
$|H'\rangle = \cos(19^{\circ})|H\rangle + \sin(19^{\circ})|V\rangle$
, $|V'\rangle = -\sin(19^{\circ})|H\rangle +
\cos(19^{\circ})|V\rangle$ with the eigenvalues $0.809$ and $0$,
respectively. The maximal violation of classicality should be
observed in the neighborhood of the state $|\phi\rangle
=\cos(67^{\circ})|H\rangle + \sin(67^{\circ})|V\rangle$.

In principle, the proposed test with the parameters obtained above
could be used to support the quantum picture for different
implementations of a qubit including, for example, "superconducting
qubits" with a still questionable  quantum character \cite{ali}.

\par
\emph{Acknowledgements.} The authors thank P. Badzi\c ag, M.
Horodecki, R. Horodecki, W.A. Majewski and M. \.Zukowski  for
discussions.   Financial support by the POLAND/SA COLLABORATION PROGRAMME of the National Research Foundation of South Africa and the Polish Ministry of Science and Higher Education and by the
European Union through the Integrated Project SCALA is acknowledged.

\end{document}